# UWF-RI2FA: Generating Multi-frame Ultrawide-field Fluorescein Angiography from Ultrawide-field Retinal Imaging Improves Diabetic Retinopathy Stratification

**Authors**

Ruoyu Chen, MD[1]†, Kezheng Xu, MD[1,4]†, Kangyan Zheng, MD[1,5]†, Weiyi Zhang, MS[1], Yan Lu, MD, PhD[6]*, Danli Shi, MD, PhD[1,2]*, Mingguang He, MD, PhD[1,2,3]*

**Affiliations**

1. Experimental Ophthalmology, School of Optometry, The Hong Kong Polytechnic University, Kowloon, Hong Kong SAR, China.

2. Research Centre for SHARP Vision (RCSV), The Hong Kong Polytechnic University, Hong Kong SAR, China

3. Centre for Eye and Vision Research (CEVR), 17W Hong Kong Science, Hong Kong SAR, China

4. State Key Laboratory of Ophthalmology, Zhongshan Ophthalmic Center, Sun Yat-sen University, Guangdong Provincial Key Laboratory of Ophthalmology and Visual Science, Guangdong Provincial Clinical Research Center for Ocular Diseases, Guangzhou, China.

5. Shantou University Medical College, Shantou, China

6. Department of Ophthalmology, The Second People's Hospital of Foshan, Affiliated Foshan Hospital of Southern Medical University, Foshan, China

† These authors contributed equally to this work

**Correspondence**

**\*Prof. Yan Lu, MD, PhD.,** The Second People's Hospital of Foshan, Affiliated Foshan Hospital of Southern Medical University, Foshan, China

Email: 837800105@qq.com

**\*Prof. Mingguang He,** MD, PhD., Chair Professor of Experimental Ophthalmology, School of Optometry, The Hong Kong Polytechnic University, Kowloon, Hong Kong SAR, China.

Email: mingguang.he@polyu.edu.hk;

**\*Dr Danli Shi,** MD, PhD., The Hong Kong Polytechnic University, Kowloon, Hong Kong SAR, China.

Email: danli.shi@polyu.edu.hk

**Abstract**

Ultrawide-field fluorescein angiography (UWF-FA) facilitates diabetic retinopathy (DR) detection by providing a clear visualization of peripheral retinal lesions. However, the intravenous dye injection with potential risks hamper its application. We aim to acquire dye-free UWF-FA images from noninvasive UWF retinal imaging (UWF-RI) using generative artificial intelligence (GenAI) and evaluate its effectiveness in DR screening. A total of 18,321 UWF-FA images of different phases were registered with corresponding UWF-RI images and fed into a generative adversarial networks (GAN)-based model for training. The quality of generated UWF-FA images was evaluated through quantitative metrics and human evaluation. The DeepDRiD dataset was used to externally assess the contribution of generated UWF-FA images to DR classification, using area under the receiver operating characteristic curve (AUROC) as outcome metrics. The generated early, mid, and late phase UWF-FA images achieved high authenticity, with multi-scale similarity scores ranging from 0.70 to 0.91 and qualitative visual scores ranging from 1.64 to 1.98 (1=real UWF-FA quality). In fifty randomly selected images, 56% to 76% of the generated images were difficult to distinguish from real images in the Turing test. Moreover, adding these generated UWF-FA images for DR classification significantly increased the AUROC from 0.869 to 0.904 compared to the baseline model using UWF-RI images ($P < .001$). The model successfully generates realistic multi-frame UWF-FA images for enhancing DR stratification without intravenous dye injection.

## Introduction

Ultra-widefield (UWF) imaging system has dramatically improved retinal lesion detection by providing 200° field of view (FOV) with abundant information on peripheral retina.[1] UWF retinal imaging (UWF-RI) and fluorescein angiography (UWF-FA) significantly increased the screening rate of lesions outside the central retina, facilitating the management of several sight-threatening diseases.[2] In diabetic retinopathy (DR), one of the leading causes of blindness,[3] peripheral lesions and larger non-perfusion area detected by UWF imaging have been demonstrated to be highly associated with disease severity[4,5] and progression risk.[6-8] These findings highlight the importance of using UWF imaging system in DR management.

Following the intravenous injection of fluorescein, UWF-FA offers a clearer visualization of vascular abnormalities and achieves more accurate DR stratification than UWF-RI.[9,10] Nevertheless, the invasiveness, various contradictions, and adverse events like vomiting, shock, potential retinal toxicity, etc.[11] make UWF-FA unsuitable as a routine screening method. Although recent advances in optical coherence tomography angiography (OCTA) have made it possible to reveal retinal circulation using noninvasive techniques, even the most advanced OCTA systems fail to provide vascular information beyond 150° FOV as effectively as UWF-FA.[12,13]

It is crucial to develop a viable solution to obtain UWF-FA images while avoiding the side effects. Recent generative artificial intelligence (GenAI) have provided practical avenues for cross-modal generation.[14,15] Generative adversarial networks (GAN) represent a popular type of generative architecture in the ophthalmic field. Although GAN have achieved outstanding performance in translating color fundus images into angiographic images and aiding in disease classification,[16,17] most of the existing explorations only focused on standard-field images with a 55° FOV. Since uncontrollable peripheral distortion, artifacts from eyelashes and lid margins are common during the process of UWF image collection,[2] cross-modal registration of UWF images is more complicated than that of standard-field fundus images, thereby hindering high-quality translation from UWF-RI to UWF-FA images. Additionally, no preliminary study aims to generate multi-frame UWF-FA images to mimic the dynamic changes of retinal circulation.

To bridge the gap, the purpose of this study is to develop a GAN-based model for generating realistic UWF-FA images of different phases from UWF-RI images on the premise of accurate cross-modal registration. The quality of translated UWF-FA images will be evaluated through quantitative parameters and clinical ophthalmologists. Moreover, the performance of translated images in aiding DR screening will be validated using an external dataset. Our UWF RI-FA solution is anticipated to provide a safe, cost-effective, and easily accessible approach for the precise management of DR in clinical scenarios.

## Methods

### Data

An inhouse dataset encompassing 3,885 UWF-RI images and 23,332 UWF-FA images of 1,352 patients from The Second People's Hospital of Foshan were utilized for model development. All UWF-RI and full-phase UWF-FA images were captured using Optos California (Optos, PLC) with a 200° field of view. The image resolution is 4000 × 4000 pixels. The private information of patients was anonymized and de-identified. Since fluorescein fully fills the entire retinal circulation only after the venous phase, the model was trained on post-venous phase images to extract the complete retinal vessel map. These post-venous UWF-FA images were divided into early-phase (25-60 seconds after fluorescein injection), mid-phase (1-5 minutes), and late-phase (after 5 minutes).

Subsequently, we used DeepDRiD as an external dataset to evaluate the performance of translated UWF-FA images in enhancing DR classification.[18] This publicly available dataset incorporates 256 UWF-RI images with labels of image quality scores and DR severity. DR was graded based on the international clinical DR classification scale (0 = no DR, 1 = mild non-proliferative DR, 2 = moderate non-proliferative DR, 3 = severe non-proliferative DR, 4 = proliferative DR).[19]

### Cross-modal registration

First, we cropped out the low-quality region of the UWF images, ie, the eyelid artifacts and iris vignetting artifacts in the periphery, then UWF-RI and UWF-FA images from the same eye and visit were registered. This pairwise registration process was performed between UWF-RI and

UWF-FA images using common retinal vessel features. Specifically, we utilized the Retina-based Microvascular Health Assessment System (RMHAS) segmentation module to extract retinal vessel maps on both image modalities and achieve pixel-to-pixel matching.[20,21] In the process, AKAZE detector was utilized to identify key points on the corresponding vessel maps,[22] and the RANSAC (random sample consensus) algorithm was employed to generate homography matrices and eliminate outliers to facilitate registration.[23] A validity restriction was integrated to ensure the accuracy of the registration, which restricted a rotation scale value between 0.8 and 1.3, and an absolute rotation radian value less than 2 before the warping transformation. Finally, we excluded image pairs with poor registration performance, like dice coefficient < 0.5, which was determined empirically based on the training dataset.

**UWF-RI to UWF-FA translation**

After quality control and preprocessing, the cross-modal image pairs were split at an 8:1:1 ratio on patient level for training, validation, and testing, respectively. The model was initialized with venous FA weight trained from 55° view FA images.[16] During training, UWF-RI images were fed into pix2pixHD,[24] a widely-used GAN model recognized for high-resolution image generation, to progressively produce realistic UWF-FA images. Images were resized to a resolution of 1024 × 1024 pixels. The paired UWF-FA images from different phases were utilized as ground truth to train three independent models for generating UWF-FA images of the early, mid, and late-phase. To enhance the generation of high-frequency components, such as retinal structure and lesions, we incorporated Gradient Variance Loss[25] to the existing loss function for a modified pix2pixHD model. Additionally, extensive data augmentations were applied during training to improve performance and prevent overfitting, including random resized crops at a scale of 0.3-3.5, random horizontal or vertical flipping, and random rotations within a range of 0-45 degrees. The models were trained with a batch size of 4 and a learning rate of 0.0002. Each training session was preset to run for 50 epochs

**Quantitative evaluation**

For the evaluation of image authenticity, we employed four standard objective measures widely used in image generation for our internal test set. Mean absolute error (MAE) computes the average absolute pixel difference between the generated and corresponding real images.[26] It quantifies the overall discrepancy in pixel values, indicating the level of fidelity in generated details. Peak signal-to-noise ratio (PSNR) approximates human perception regarding reconstruction quality. It measures the ratio between the maximum possible power of a signal and the power of the noise interfering with it.[27] Structural similarity measures (SSIM) assess the structural similarity between images, with 1 representing complete similarity and 0 indicating no similarity.[28] SSIM provides insights into the visual resemblance and coherence between the generated and real images. Multi-scale SSIM (MS-SSIM) provides more flexibility in incorporating variations in viewing conditions and image resolution. The higher the SSIM, MS-SSIM, and PSNR, the better the quality of the generated images.

**Human evaluation**

**Visual quality evaluation:** a subset consisting of fifty pairs encompassing of real UWF-RI, real and generated UWF-FA images from the test set was randomly selected and assigned to two experienced ophthalmologists (K.Z and K.X). The evaluation process was conducted using a standardized five-tier scale, considering the overall similarity between real and generated UWF-FA images, the integrity of the retinal structure and lesions, as well as the rationality of fluorescein alterations in different phases (1=the quality of the generated UWF-FA images is equivalent to that of the corresponding real ones, 5=extremely poor quality). Detailed grading criteria and illustrative examples of varying image qualities were compiled and presented in Supplementary Figure 1 and 2. The inter-grader agreement was quantitatively determined using Cohen's linearly weighted kappa, which ranges from -1 (complete disagreement) to 1 (perfect agreement). In practice, scores within the ranges of 0.4 to 0.6, 0.6 to 0.8, and 0.8 to 1.0 indicate moderate, substantial, and almost perfect agreement, respectively.

**Turing test:** We then conducted a Turing test. Specifically, we asked two ophthalmologists (K.Z and K.X) to determine whether the UWF-FA images were (1) real images acquired from patients or (2) generated images using our GAN-based model. For each category, 25 images were included.

**Applicability evaluation**

We leveraged the DeepDRiD dataset to assess the effectiveness of these synthetic UWF-FA images for DR severity grading. The real UWF-RI images from DeepDRiD dataset were utilized as input for generating corresponding UWF-FA images of early, mid and late-phase. The dataset was divided into training, validation, and test sets with a ratio of 3:1:1. During training, the images were resized to 1024×1024 pixels and augmented with random horizontal flips and rotations between -30 and 30 degrees. The Adam optimizer was used with a learning rate of 1e-5 and a batch size of 4. The Swin Transformer and multi-layer perceptron were employed as a feature extractor and classifier to investigate whether integrating generated UWF-FA images could improve DR stratification. The same hyperparameters were used in each experiment to classify DR based on (1) real UWF-RI images (group 1); (2) real UWF- RI and generated early phase UWF-FA images (group 2); (3) real UWF- RI, generated early and mid-phase UWF-FA images (group 3); (4) real UWF- RI, generated early, mid and late-phase UWF-FA images (group 4). These four models were initialized with pretrained weights from ImageNet and shared the same training data.[29] The Swin Transformer extracted features from different images into 1024-dimensional embeddings.[30] These embeddings were then concatenated, passed through a fully connected layer, and followed by a softmax layer to obtain the classification results. The performance of these models was compared using the area under the receiver operating characteristic curve (AUROC), F1 scores, and accuracy metrics.

## Results

The images of poor quality and those that could not be matched pairwisely were excluded. The final dataset consisted of 2,747 UWF-RI images and 18,321 UWF-FA images from 1,263 patients. An average of 2 UWF-RI images and 15 UWF-FA images were obtained from each patient. The median (inter-quartile range, IQR) age of patients was 55.19 (43.50 to 65.71) years, and 722 (57.2%) patients were male. Each UWF-FA image was matched with its corresponding UWF-RI image, captured from the same eye and during the same visit. We finally yielded 18,321 UWF RI-FA pairs for model development, of which 1,863 pairs were in the early-phase, 10,275 pairs were in the mid-phase, and 6,183 pairs were in the late-phase (Figure 1).

**Quantitative evaluation**

Pixel-wise similarity between real UWF-FA images and those generated from UWF-RI images was assessed on internal test set. For the generated early-phase UWF-FA images, the MAE, PSNR, SSIM, and MS-SSIM were 100.19, 28.20, 0.83, and 0.89. For the generated mid-phase UWF-FA images, the MAE, PSNR, SSIM, and MS-SSIM were 106.48, 27.49, 0.84, and 0.88. For the generated late-phase UWF-FA images, the MAE, PSNR, SSIM, and MS-SSIM were 113.90, 29.09, 0.84, and 0.91. These results are shown in Table 1.

**Human evaluation**

**Visual quality evaluation:** The visual quality evaluation was conducted by two human experts on a five-tier scale using fifty sets of real UWF-RI images, as well as real and generated UWF-FA images. These results are presented in Table 2. For early-phase UWF-FA images, raters gave mean (SD) grades of 1.88 (1.00) and 1.92 (1.13), respectively, with an inter-rater kappa of 0.81. For mid-phase images, the grades were 1.86 (0.87) and 1.98 (0.97), yielding a kappa of 0.79. The late-phase images were rated with 1.74 (0.63) and 1.64 (0.69), achieving a kappa coefficient of 0.84.

**Turing test:** Two ophthalmologists correctly identified 80%-88% of the real UWF-FA images. For images generated using the proposed model, two ophthalmologists mistook generated UWF-FA images as real images in 56%-76% of the cases. These results from human evaluation suggest that most of the synthesized images achieved high authenticity.

**Performance on aiding DR classification**

Performance metrics of generated UWF-FA images in aiding DR classification using the DeepDRiD dataset are shown in Table 3. Classifier trained on merely UWF-RI images was taken as comparative baseline and got an AUROC of 0.869, an AUPR of 0.717, and an F1 score of 0.599. In contrast, a classifier trained on a combination of UWF-RI and generated early-phase UWF-FA images improved the AUROC to 0.886 (P = .015), with further improvement to 0.887 (P < .001) when generated mid-phase UWF-FA images were included. Using UWF-RI and generated early, mid, and late-phase UWF-FA images, the best classifier achieved an AUROC of 0.904 (P<.001), an AUPR of 0.785, and an F1 score of 0.625. All the metrics of classification performance were significantly improved from the comparative baseline.

## Discussion

The current study pioneeringly achieved accurate cross-modal registration of large-scale UWF images and proposed a GAN-based approach for obtaining dye-free multi-frame UWF-FA images from noninvasive UWF-RI images. The UWF RI-FA translation model has been validated through multiscale quantitative metrics and human evaluation for its ability to generate high-quality UWF-FA images with realistic retinal structures and lesions. Most importantly, we achieved a significant enhancement in DR grading by integrating our translated UWF-FA images into an external dataset. The current study offers a promising non-invasive method to bypass fluorescein injection and improve DR screening.

GenAI is a promising technology that has been applied in ophthalmic field[31-35] and GAN is a representative configuration.[36] We have applied GAN to generate 55° view FA and ICGA images from color fundus images.[16,17] Here, we are making a step forward to generate 200° multi-frame UWF-FA images from UWF-RI images. Generating realistic vascular lesions on UWF-FA images is difficult due to the relatively inconspicuous manifestation on UWF-RI images.[37] Thus, we took advantage of transfer learning with an established GAN-based backbone pretrained on a large-scale dataset of standard field RI-FA pairs to capture the association between lesions on color fundus images and FFA.[16] Moreover, we made a crucial modification by adjusting the loss function of our model, adding Gradient Variance Loss to the original adversarial loss and pixel-reconstruction loss.[25] This modification contributed to generate realistic high-resolution details on UWF-FA images.

Previously, Fang et al proposed UWAT-GAN to synthesize venous UWF-FA image from UWF-RI image.[38] This study applied a relatively small dataset consisting of 70 paired images and found it difficult to achieve satisfactory registration because of the uncontrollable artifacts during image collection. Additionally, the UWAT-GAN generated only a single UWF-FA image from a UWF-RI image, without accounting for the dynamic changes in retinal circulation and lesions during the UWF-FA process. In the current study, our model could generate realistic UWF-FA images of early-phase, mid-phase, and late-phase. These multi-frame images reflect lesion features in a specific time frame, providing additional information of retinal abnormalities like real UWF-FA examination. Moreover, apart from objective evaluation metrics, the current study also asked experienced ophthalmologists to perform visual score assessments and a Turing test. The results validated the authenticity of the generated UWF-FA images.

UWF imaging system integrates laser ophthalmoscope technology with unique optical properties of ellipsoidal mirror, enabling it to capture approximately 80% of the entire retinal area.[1] Previous studies demonstrated that the UWF imaging system nearly doubled the detection rate of DR. Additionally, the identification of peripheral lesions indicated a more severe level of DR in nearly 10% of eyes, suggesting the importance of detecting peripheral lesions in DR screening.[4] UWF-FA serves as the gold standard examination in capturing vascular abnormalities. Nevertheless, the complex collection process and several side effects related to fluorescein injection hinder its use in routine clinical settings. Cross-modal translation for UWF images using GenAI can incorporate advantages of the accessible image modality and provide potential alternatives to invasive imaging.[39] The proposed GAN-based model shows strong power in translating noninvasive UWF-RI to realistic UWF-FA images, avoiding several risks related to intravenous fluorescein injection. The results also illustrated that integrating full-phase generated UWF-FA images significantly enhance DR stratification in downstream task using an external dataset. Our UWF RI-FA model offers a potential approach for obtaining dye-free UWF-FA images and enhancing multimodal imaging detection of retinal abnormalities.

Some inevitable limitations should be noted. Although accurate registration between paired UWF-RI and UWF-FA images was achieved using vascular features, the peripheral artifacts and distortion still interfered the generation quality of peripheral lesions. High-quality UWF RI-FA pairs are needed for model optimization. Moreover, this study was conducted in the research setting, and thus, real-world validation is needed to further assess the generalizability and robustness of the UWF RI-FA cross-modal translation. Also, the generated UWF-FA images were merely applied in the DR screening task, while UWF imaging has a wide application in retinal diseases.[40] Abundant and diverse data is required to extend the applicable scope of generated UWF-FA images.

In conclusion, our study introduces a paradigm-shifting GAN-based approach for translating UWF-RI images into multi-frame UWF-FA images, marking a potential advancement in non-invasive retinal disease screening. The rigorous training and validation of our modality-translation model on a large-scale dataset have yielded high-quality UWF-FA images that accurately depict critical retinal features and lesions. Integration of these translated images has led to a notable improvement in DR grading performance, demonstrating the model's potential to enhance DR screening and management. This innovation not only addresses the need for a safer and more accessible diagnostic tool but also contributes to the broader application of GenAI in improving ophthalmic care.

# Reference

1. Vujosevic S, Aldington SJ, Silva P, et al. Screening for diabetic retinopathy: new perspectives and challenges. *The Lancet Diabetes & Endocrinology* 2020; **8**(4): 337-47.
2. Kumar V, Surve A, Kumawat D, et al. Ultra-wide field retinal imaging: A wider clinical perspective. *Indian Journal of Ophthalmology* 2021; **69**(4): 824.
3. Ockrim Z, Yorston D. Managing diabetic retinopathy. *BMJ* 2010; **341**(oct25 1): c5400-c.
4. Silva PS, Cavallerano JD, Sun JK, Soliman AZ, Aiello LM, Aiello LP. Peripheral Lesions Identified by Mydriatic Ultrawide Field Imaging: Distribution and Potential Impact on Diabetic Retinopathy Severity. *Ophthalmology* 2013; **120**(12): 2587-95.
5. Zhuang X, Chen R, Liang A, et al. Multimodal imaging analysis for the impact of retinal peripheral lesions on central neurovascular structure and retinal function in type 2 diabetes with diabetic retinopathy. *British Journal of Ophthalmology* 2023; **107**(10): 1496-501.
6. Silva PS, Horton MB, Clary D, et al. Identification of Diabetic Retinopathy and Ungradable Image Rate with Ultrawide Field Imaging in a National Teleophthalmology Program. *Ophthalmology* 2016; **123**(6): 1360-7.
7. Silva PS, Marcus DM, Liu D, et al. Association of Ultra-Widefield Fluorescein Angiography–Identified Retinal Nonperfusion and the Risk of Diabetic Retinopathy Worsening Over Time. *JAMA Ophthalmology* 2022; **140**(10): 936.
8. Silva PS, Cavallerano JD, Haddad NMN, et al. Peripheral Lesions Identified on Ultrawide Field Imaging Predict Increased Risk of Diabetic Retinopathy Progression over 4 Years. *Ophthalmology* 2015; **122**(5): 949-56.
9. Ashraf M, AbdelAl O, Shokrollahi S, Pitoc CM, Aiello LP, Silva PS. Evaluation of diabetic retinopathy severity on ultrawide field colour images compared with ultrawide fluorescein angiograms. *British Journal of Ophthalmology* 2023; **107**(4): 534-9.
10. Ashraf M, Sampani K, AbdelAl O, et al. Disparity of microaneurysm count between ultrawide field colour imaging and ultrawide field fluorescein angiography in eyes with diabetic retinopathy. *British Journal of Ophthalmology* 2020; **104**(12): 1762-7.
11. Kornblau IS, El-Annan JF. Adverse reactions to fluorescein angiography: A comprehensive review of the literature. *Surv Ophthalmol* 2019; **64**(5): 679-93.
12. Niederleithner M, De Sisternes L, Stino H, et al. Ultra-Widefield OCT Angiography. *IEEE Transactions on Medical Imaging* 2023; **42**(4): 1009-20.
13. Ni S, Liang GB, Ng R, et al. Panretinal handheld OCT angiography for pediatric retinal imaging. *Biomed Opt Express* 2024; **15**(5): 3412-24.
14. Goodfellow IJ, Pouget-Abadie J, Mirza M, et al. Generative Adversarial Networks. 2014.
15. You A, Kim JK, Ryu IH, Yoo TK. Application of generative adversarial networks (GAN) for ophthalmology image domains: a survey. *Eye and Vision* 2022; **9**(1): 6.
16. Shi D, Zhang W, He S, et al. Translation of Color Fundus Photography into Fluorescein Angiography Using Deep Learning for Enhanced Diabetic Retinopathy Screening. *Ophthalmology Science* 2023; **3**(4): 100401.
17. Chen R, Zhang W, Song F, et al. Translating color fundus photography to indocyanine green angiography using deep-learning for age-related macular degeneration screening. *npj Digital Medicine* 2024; **7**(1): 34.
18. Liu R, Wang X, Wu Q, et al. DeepDRiD: Diabetic Retinopathy—Grading and Image Quality Estimation Challenge. *Patterns* 2022; **3**(6): 100512.

19. Grading Diabetic Retinopathy from Stereoscopic Color Fundus Photographs — An Extension of the Modified Airlie House Classification. *Ophthalmology* 2020; **127**(4): S99-S119.

20. Shi D, Lin Z, Wang W, et al. A Deep Learning System for Fully Automated Retinal Vessel Measurement in High Throughput Image Analysis. *Frontiers in Cardiovascular Medicine* 2022; **9**: 823436.

21. Shi D, He S, Yang J, Zheng Y, He M. One-shot Retinal Artery and Vein Segmentation via Cross-modality Pretraining. *Ophthalmology Science* 2024; **4**(2): 100363.

22. Alcantarilla PF, Nuevo J, Bartoli A. Fast Explicit Diffusion for Accelerated Features in Nonlinear Scale Spaces.  British Machine Vision Conference; 2013; 2013.

23. Cantzler H. Random Sample Consensus ( RANSAC ).

24. Isola P, Zhu J-Y, Zhou T, Efros AA. Image-to-Image Translation with Conditional Adversarial Networks. *2017 IEEE Conference on Computer Vision and Pattern Recognition (CVPR)* 2016: 5967-76.

25. Abrahamyan L, Truong AM, Philips W, Deligiannis N. Gradient Variance Loss for Structure-Enhanced Image Super-Resolution.  ICASSP 2022 - 2022 IEEE International Conference on Acoustics, Speech and Signal Processing (ICASSP); 2022 2022-5-23: IEEE; 2022. p. 3219-23.

26. Chourak H, Barateau A, Tahri S, et al. Quality assurance for MRI-only radiation therapy: A voxel-wise population-based methodology for image and dose assessment of synthetic CT generation methods. *Frontiers in Oncology* 2022; **12**: 968689.

27. Tanabe Y, Ishida T. Quantification of the accuracy limits of image registration using peak signal-to-noise ratio. *Radiological Physics and Technology* 2017; **10**(1): 91-4.

28. Wang Z, Bovik AC, Sheikh HR, Simoncelli EP. Image Quality Assessment: From Error Visibility to Structural Similarity. *IEEE Transactions on Image Processing* 2004; **13**(4): 600-12.

29. Russakovsky O, Deng J, Su H, et al. ImageNet Large Scale Visual Recognition Challenge. *International Journal of Computer Vision* 2014; **115**: 211 - 52.

30. Liu Z, Lin Y, Cao Y, et al. Swin Transformer: Hierarchical Vision Transformer using Shifted Windows. *2021 IEEE/CVF International Conference on Computer Vision (ICCV)* 2021: 9992-10002.

31. He S, Joseph S, Bulloch G, et al. Bridging the Camera Domain Gap With Image-to-Image Translation Improves Glaucoma Diagnosis. *Transl Vis Sci Technol* 2023; **12**(12): 20.

32. Chen X, Zhang W, Xu P, et al. FFA-GPT: an automated pipeline for fundus fluorescein angiography interpretation and question-answer. *NPJ Digit Med* 2024; **7**(1): 111.

33. Chen X, Zhang W, Zhao Z, et al. ICGA-GPT: report generation and question answering for indocyanine green angiography images. *Br J Ophthalmol* 2024.

34. Waisberg E, Ong J, Kamran SA, et al. Generative artificial intelligence in ophthalmology. *Surv Ophthalmol* 2024.

35. Shi D, Zhang W, Chen X, et al. EyeFound: A Multimodal Generalist Foundation Model for Ophthalmic Imaging. *ArXiv* 2024; **abs/2405.11338**.

36. Wang Z, Lim G, Ng WY, et al. Generative adversarial networks in ophthalmology: what are these and how can they be used? *Curr Opin Ophthalmol* 2021; **32**(5): 459-67.

37. Marcus DM, Silva PS, Liu D, et al. Association of Predominantly Peripheral Lesions on Ultra-Widefield Imaging and the Risk of Diabetic Retinopathy Worsening Over Time. *JAMA Ophthalmol* 2022; **140**(10): 946-54.

38. Fang Z, Chen Z, Wei P, et al. UWAT-GAN: Fundus Fluorescein Angiography Synthesis via Ultra-wide-angle Transformation Multi-scale GAN. *ArXiv* 2023; **abs/2307.11530**.

39. Lyu J, Fu Y, Yang M, et al. Generative Adversarial Network-based Noncontrast CT Angiography

for Aorta and Carotid Arteries. *Radiology* 2023; **309**(2): e230681.

40. Kumar V, Surve A, Kumawat D, et al. Ultra-wide field retinal imaging: A wider clinical perspective. *Indian J Ophthalmol* 2021; **69**(4): 824-35.

# Acknowledgement

**Funding**

Danli Shi and Mingguang He disclose support for the research and publication of this work from the Start-up Fund for RAPs under the Strategic Hiring Scheme (Grant Number: P0048623) and the Global STEM Professorship Scheme (Grant Number: P0046113) from HKSAR. The funders had no role in study design, data collection and analysis, decision to publish, or manuscript preparation.

**Author contributions**

D.S. and M.H. conceived the study. D.S. built the deep learning model. D.S., W.Z. and R.C. analyzed the data. M.H., and Y.L. provided data. R.C., K.X. and K.Z contributed to literature review and key data interpretation. R.C., K.X. and K.Z wrote the manuscript. All authors critically revised the manuscript.

**Conflicts of interest**

Mingguang He and Danli Shi. are inventors of the technology mentioned in the study patented as "A method to translate fundus photography to realistic angiography" (CN115272255A). Other authors declare no conflict of interest.".

**Data and code availability**

The data used for model development of this study are not openly available due to reasons of privacy. The authors do not have the permission to distribute the dataset publicly. Code is available at: https://github.com/NVIDIA/pix2pixHD

**Ethics**

We utilized de-identified existing data for our study, which received approval from the Institutional Review Board of the Hong Kong Polytechnic University.

# Figures

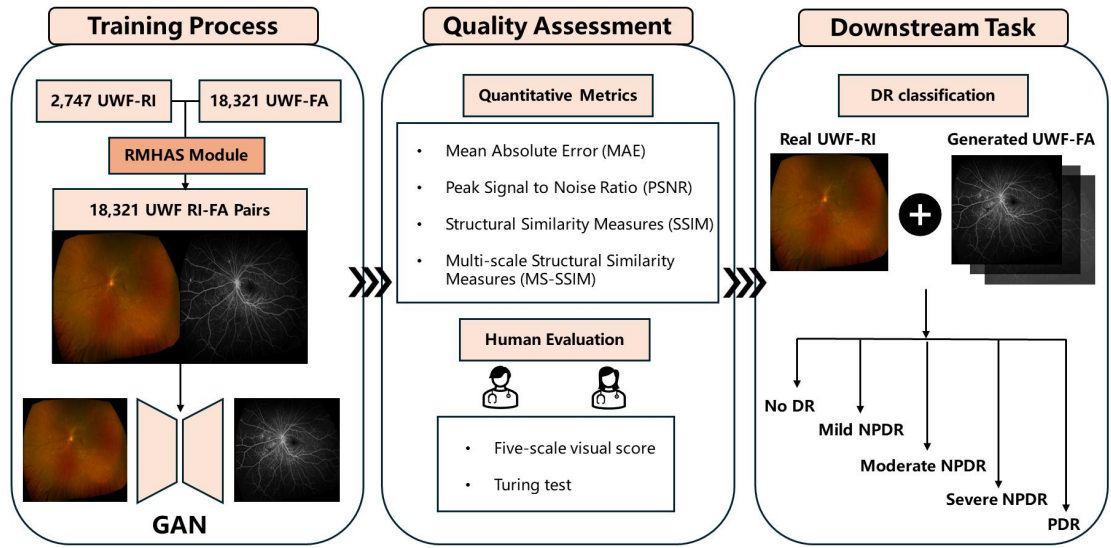

**Fig.1. Flowchart of this study.** GAN=generative adversarial networks, UWF-RI= ultrawide-field retinal imaging, UWF-FA=ultrawide-field fundus fluorescein angiography, RMHAS=Retina-based Microvascular Health Assessment System, DR=diabetic retinopathy, NPDR=nonproliferative diabetic retinopathy, PDR=proliferative diabetic retinopathy

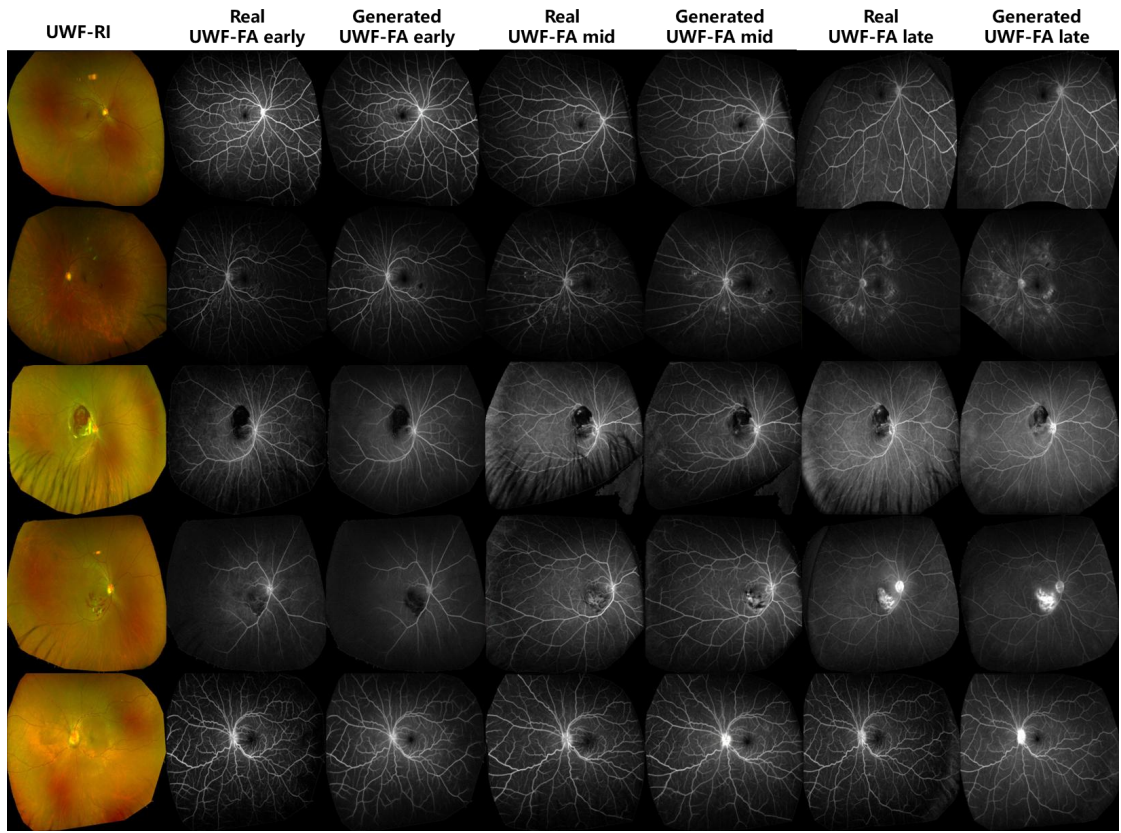

**Fig 2. Examples of real and generated ultrawide-field fundus fluorescein angiography (UWF-FA) images in the early, mid, and late phases.** UWF-RI= ultrawide-field retinal imaging, UWF-FA=ultrawide-field fundus fluorescein angiography. 1st Row: Normal; 2nd Row: Diabeitic Retinopathy; 3rd Row: Retinal Macroaneurysms; 4th Row: Retinal Vein Occlusion; 5th Row: Papilloedema

**Tables**

**Table 1.** Quantitative evaluation results of generated ultrawide-field fluorescein angiography (UWF-FA) images

|  | MAE | PSNR | SSIM | MS-SSIM |
|---|---|---|---|---|
| **Early-phase UWF-FA** | 113.90 (±53.78) | 29.09 (±3.72) | 0.84 (±0.06) | 0.91 (±0.04) |
| **Mid-phase UWF-FA** | 100.19 (±44.94) | 28.20 (±3.02) | 0.83 (±0.04) | 0.70 (±0.04) |
| **Late-phase UWF-FA** | 106.48 (±45.93) | 27.49 (±4.25) | 0.84 (±0.06) | 0.88 (±0.07) |

MAE=mean absolute error, PSNR=peak signal-to-noise ratio, SSIM=structural similarity measures, MS-SSIM=multi-scale structural similarity measures, UWF-FA=ultrawide-field fluorescein angiography

**Table 2.** Visual quality assessment of generated ultra-widefield fluorescein angiography (UWF-FA) in an internal test set (N=50) using a five-point visual scale.

|  | Rater 1 [Mean (SD)] | Rater 2 [Mean (SD)] | Kappa |
|---|---|---|---|
| **Early-phase UWF-FA** | 1.88 (1.00) | 1.92 (1.13) | 0.81 |
| **Mid-phase UWF-FA** | 1.86 (0.87) | 1.98 (0.97) | 0.79 |
| **Late-phase UWF-FA** | 1.74 (0.63) | 1.64 (0.69) | 0.84 |

UWF-FA =ultrawide-field fluorescein angiography, SD= standard deviation.

**Table 3.** Comparison of using a real ultrawide-field retinal imaging (UWF-RI) alone versus the integration of generated ultrawide-field fluorescein angiography (UWF-FA) in diabetic retinopathy (DR) stratification on an external dataset (DeepDRiD, N=250)

| Group | AUROC | AUPR | F1-score | Sensitivity | Specificity | Accuracy | P value |
|---|---|---|---|---|---|---|---|
| **Group 1** | 0.869 | 0.717 | 0.599 | 0.601 | 0.859 | 0.784 | |
| **Group 2** | 0.886 | 0.756 | 0.614 | 0.639 | 0.869 | 0.796 | 0.015* |
| **Group 3** | 0.887 | 0.787 | 0.625 | 0.649 | 0.882 | 0.821 | <0.001* |
| **Group 4** | 0.904 | 0.785 | 0.625 | 0.658 | 0.886 | 0.833 | <0.001* |

AUROC=area under the receiver operating characteristic curve; AUPR=area under the precision-recall curve; 1= only using real UWF retinal imaging (UWF-RI) for DR classification; 2=combining real UWF-RI images and generated early UWF-FA images for DR classification; 3= combining real UWF-RI images and generated early+mid UWF-FA images for DR classification; 4= combining real UWF-RI images and generated early+mid+late UWF-FA images for DR classification; *P < 0.05.

**List of Supplementary Material**

Supplementary Fig.1. Demonstration of the grading criteria in human evaluation

Supplementary Fig. 2. Generated UWF-FA images of different visual quality.